\documentclass[
 floatfix,twocolumn,superscriptaddress, prl,aps
 amsmath,amssymb,
 aps,
]{revtex4-2}
\usepackage[utf8]{inputenc}
\usepackage{newunicodechar}
\newunicodechar{，}{,}
\usepackage{amsmath}
\usepackage{amsfonts}
\usepackage{amssymb,amsthm}

\usepackage[normalem]{ulem}
\usepackage{amsmath}
\newcommand{\stkout}[1]{\ifmmode\text{\sout{\ensuremath{#1}}}\else\sout{#1}\fi}
\usepackage[subrefformat=parens,labelformat=parens,caption=false]{subfig}
\usepackage{graphicx}
\usepackage{censor}
\usepackage{multirow}
\usepackage[]{hyperref}
\usepackage{bbold}
\usepackage{cancel}
\hypersetup{
  colorlinks = true,
  urlcolor = magenta,
  linkcolor = red,
  citecolor = blue
}
\usepackage[normalem]{ulem}
\usepackage{braket}
\usepackage[all]{hypcap}
\usepackage{url}
\usepackage{float}

\usepackage{xcolor}

\usepackage[]{hyperref}
\usepackage{bbold}
\usepackage{bm}
\usepackage{cancel}
\newcommand{\eq}[1]{\begin{align}#1\end{align}}

\usepackage{graphicx}
\graphicspath {{figures/}}

\begin{document}
\title{Artificial-atom arrays in moiré superlattices for quantum optics  
}
\author{Zhigang Song}
\altaffiliation{These authors contributed equally to this work} 
\affiliation{Center for Quantum Matter, School of Physics, Zhejiang University, Hangzhou 310058, China}
\affiliation{Contact author: zsong@zju.edu.cn or kchang@zju.edu.cn}

\author{Peng Xu}
\altaffiliation{These authors contributed equally to this work}
\affiliation{Quantum Information Institute, School of Physics, Zhengzhou University, Zhengzhou 450001, China}
\affiliation{Institute of Quantum Materials and Physics, Henan Academy of Sciences, Zhengzhou 450046, China}

\author{Kai Chang}
\affiliation{Center for Quantum Matter, School of Physics, Zhejiang University, Hangzhou 310058, China}
\affiliation{Institute for Advanced Study in Physics, Zhejiang University, Hangzhou 310058, China}
\affiliation{Contact author: zsong@zju.edu.cn or kchang@zju.edu.cn}

\begin{abstract}
Solid-state platforms are particularly attractive for quantum optics because they facilitate on-chip integration and are compatible with established semiconductor and photonic technologies. However, a major challenge in solid-state quantum optics is the fabrication of arrays of identical emitters, such as quantum dots. In this work, we propose moiré superlattices as a novel solid-state platform for manipulating light at the single-photon level. Moiré superlattices form arrays of artificial-atom states characterized by nearly identical optical transition energies, tunable spacing, and highly adjustable electronic structures. They naturally operate as atomically thin, scalable, periodic emitters, making them ideal for quantum applications. Additionally, the extensive materials database of moiré superlattices offers spectral coverage spanning a broad range of optical wavelengths.

\end{abstract}
\maketitle



\emph{Introduction---} Controlling light at the single-photon level is a central goal in modern quantum science~\cite{scully1997quantum, vogel2006quantum, PRXQuantum.5.010344, sun2018single}, as it underpins a wide range of emerging technologies in quantum communication~\cite{ding2025high, zheng2026large}, computation~\cite{Justin2013Boson, Oh2025Computer}, and sensing~\cite{RevModPhys.89.035002}. Since direct photon-photon interactions do not occur in vacuum, the control of photons has to be done via a light-matter interface, where photons are coupled to emitters possessing discrete electronic energy levels\cite{chang2018colloquium}. An effective light-matter interface requires that the emitters are identical and have the same optical properties in an array~\cite{sierra2022dicke, RevModPhys.82.209}.

   
Although neutral alkali atoms are among the most promising platforms for quantum optics~\cite{saffman2010quantum, Scully2015Single, rui2020subradiant, Zhang2020Subradiant, Patti2021Controlling, meng2023atomic}, solid-state light–matter interfaces are particularly attractive owing to their potential for on-chip integration and compatibility with established semiconductor and photonic technologies. Various solid-state systems have been developed to approach the single-photon regime, including quantum dots coupled to nanocavities~\cite{reitzenstein2012semiconductor}, color centers in diamond~\cite{castelletto2020silicon}, superconducting circuits interacting with microwave photons~\cite{blais2020quantum}, confined electrons and defects in solid states. However, a major challenge in all these systems is the difficulty of fabricating identical emitters, particularly within solid-state platforms. Developing scalable, solid-state arrays of identical emitters remains a long-sought goal for quantum optics~\cite{moreno2021quantum, ebadi2021quantum, shaw2026cavity}.

Recently, moiré superlattices have emerged as exciting concept in condensed matter physics, largely driven by the study of strongly-correlated phenomena\cite{kennes2020one, devakul2021magic, bi2021excitonic}. Moiré superlattices are formed by stacking two layered materials with a relative twist, resulting in a larger-scale periodic pattern and the emergence of flat bands~\cite{xian2021realization, an2021moire, angeli2021gamma, PhysRevLett.125.037402, soltero2022moire, PhysRevMaterials.5.014007, xian2018multi, li2021lattice, zhang2021one}. As the twist angle decreases, the bandwidth of these bands typically narrows. When the bandwidth becomes vanishingly small, the suppression of electron kinetic energy makes moiré superlattices less important for strongly correlated quantum transport, but it opens a promising avenue for quantum information applications~\cite{song2025twisted, xu2022tunable,li2022tunable, tang2020simulation, nowakowski2025single}. Besides, earlier investigations have identified nearly dispersionless bands at small twist angles~\cite{song2022deep,song2025twisted}.

In this work, we propose moiré superlattices as a novel solid-state platform for controlling light at the single-photon level. The artificial-atom states and highly tunable electronic structures of twisted bilayer materials naturally form periodic emitter arrays with nearly identical optical transition energies and controllable spacing—properties rarely achieved in conventional solid-state systems such as quantum dots or defects. Consequently, moiré superlattices offer an atomically thin, scalable, solid-state platform for quantum optics and related applications. Furthermore, the extensive material library of moiré superlattices enables spectral operation across a broad range of optical wavelengths.

\emph{Electronic structure of twisted bilayer materials---} Density functional theory (DFT) is a well-established numerical method for calculating the electronic structure of materials based on quantum mechanics. As an illustrative example, we employ DFT to investigate the band structure of 1.09°-twisted bilayer hexagonal boron nitride (h-BN) and assess its potential for applications in quantum optics. When two layered materials are stacked with an appropriate twist angle, a long-range moiré superlattice emerges. After structural relaxation, the material exhibits distinct stacking regions insides a unit cell: the AA-stacking region expands into a large triangular area, the AB-stacking region shrinks to a small portion, and the rhomboid-stacking region forms a small triangular area, as shown in Fig. S1. For a triangular moiré lattice, the lattice constant of the superlattice is given by $L=\frac{a}{2sin(\vartheta/2)}$, where $\vartheta$ is the twist angle. In twisted bilayer BN, $a=2.50 \, \text{\AA}$. In experiments, the twist angle of h-BN moiré superlattices ranges from 0.01° to 60°, corresponding to moiré lattice constants ranging from 1432 nm to 0.25 nm~\cite{hong2026highly}.

\begin{figure}
\centering
\includegraphics[width=1\linewidth]{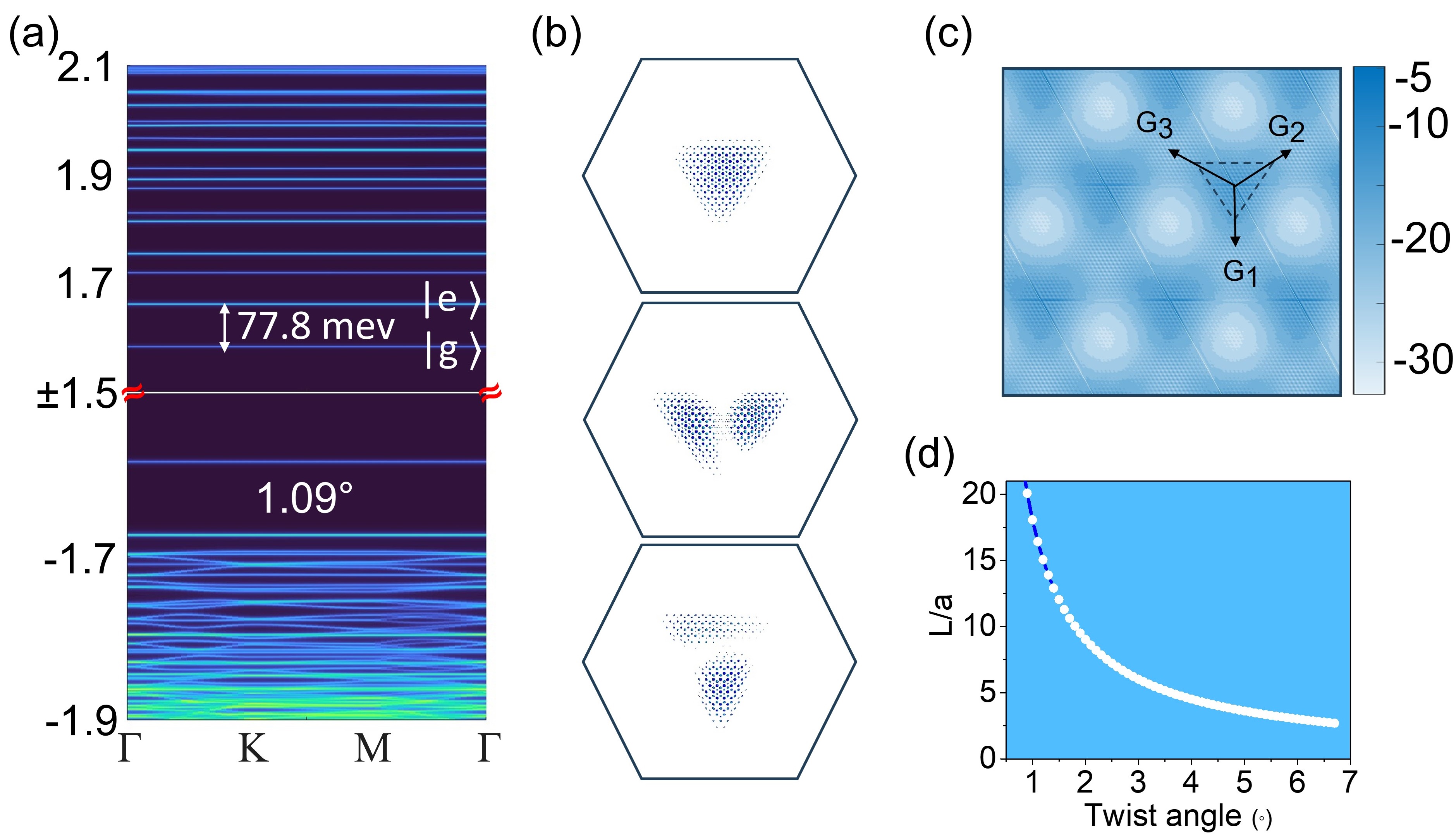}
\caption{
(a) Electronic structure of bilayer h-BN with a twist angle of 1.09°. The y-axis is truncated in the range from -1.5 to 1.5 eV, where there is no state. (b) Wave functions of three lowest bands above the Fermi level. (c) DFT-calculated onsite potential of $p_z$ orbitals, where the triangular shape illustrates the effective quantum well (in meV). (d) Moiré superlattice constant ($L$) as a function of twist angle.}
\label{fig:bansd structure}
\end{figure}

The energy bands of twisted h-BN (i.e., the energy dispersion as a function of momentum) become progressively flatter as the twist angle decreases. The band structures at different twist angles are shown in Fig. 1 and Fig. S2. When the twist angle falls below 5.09°, flat bands emerge above the Fermi level and are well separated from the other bands. As the twist angle decreases further, additional flat bands appear above the Fermi level (see Fig. S2). At a twist angle of 1.69°, the calculated bandwidth of the flat band above the Fermi level is less than 0.01 meV, which can be considered negligible given the possible error in DFT calculations. This indicates strong electron localization in real space and weak decoherence. Therefore, this angle can be regarded as a “magic angle” for quantum information applications. The corresponding band is essentially dispersionless.

The calculated band structure at a twist angle of 1.09° is shown in Fig. 1a. The dispersionless bands exhibit wavefunctions resembling those of a triangular quantum well, indicating strongly localized states (see Fig. 1b). The energy separation between the two lowest localized states above the Fermi level is approximately 77.8 meV. The second and third bands are degenerate in energy. Moreover, the energy separation between the lowest localized state and the lowest dispersive band (with a bandwidth greater than 1 meV) is as large as $\Delta b \approx 500$ meV. The DFT-calculated energy levels and wavefunctions respect the $D_3$ point-group symmetry, and their corresponding irreducible representations are presented in Fig. S3. The spatial extent of the localized states is on the order of $20 a_0$, where $a_0$ is the Bohr radius, making them roughly an order of magnitude larger than typical alkali atoms. As a result, the transition dipole matrix elements are significantly enhanced, reaching values on the order of $20 a_0 e$.

The dispersionless bands in these twisted semiconductors can be understood in a straightforward manner. The band edges of most monolayer materials can be described by an effective-mass approximation, and the moiré potential forms an ordered lattice of quantum wells. For a triangular supperlattice, the effective Hamiltonian is as follows:
\begin{align}
   H = \frac{\hbar^2 \nabla^2}{2m_{\text{eff}}}+\sum_{i=1}^3 V_0(\cos(\bm{G}_i\cdot\bm{r})-\sin(\bm{G}_i\cdot\bm{r})),
\label{eq:Hp_bright_rev}
\end{align}
where $m_{\text{eff}}$ is the effective mass of the original monolayers. In h-BN, $m_{\text{eff}} \approx 0.77 m_e$ with the electron mass $m_e$. The second term is the moiré potential with $\bm{G}_i$ being reciprocal vectors as shown in Fig. 1c. The magnitudes are given by $G_i=2\pi/L$, where $L$ the moiré superlattice constant can be tuned by the twist angle as described in Fig. 1d. The parameter $V_0$ depends on both the material and the twist angle. For a twist angle of 1.09°, a fit to the DFT results yields $V_0 = 160$ meV. The DFT-calculated onsite potential of the $p_z$ orbitals of B and N atoms in twisted bilayer h-BN is displayed in Fig. 1c. The numerical results and wavefunction symmetries obtained from Eq. 1 are in good agreement with the DFT calculations (see Fig. S4), particularly in terms of symmetry and the structure of the energy levels. The dispersionless band states in twisted bilayer materials resemble the states in arrays of alkali atoms, and thus localized states in twisted bilayer materials can be treated as arrays of artificial atoms as illustrated in Fig. 2a.

For an infinite triangular quantum well, the energy levels are given by $E=E_0(m^2-mn+n^2)$, where $m$ and $n$ are positive integer quantum numbers satisfying $m \ge 2n$. When $m>2n$, the dispersionless bands are twofold degenerate, whereas for $m=2n$, the levels are nondegenerate. The characteristic energy scale $E_0$ is approximately 16 meV. Despite its simplicity, this model captures the essential features of the system, including the energy spectrum, degeneracies, irreducible representations, and even the optical selection rules. In particular, symmetry and angular momentum impose selection rules for dipole transitions between localized states. Because the actual potential deviates from an ideal triangular shape, the predicted energy levels exhibit some quantitative deviations from DFT calculations, as compared in Fig. 2b.

Any two dispersionless bands above the Fermi level can be treated as a two-level system. Here, we focus on the two lowest such bands. When one electron is injected into each moiré supercell, an effective array of identical emitters naturally emerges. Neglecting electron spin, the Hamiltonian of this two-level system is given by (see Supporting Information for a detailed derivation):
\begin{align}
    H =\sum_{i，m} \varepsilon_{m} c^{\dagger}_{im}c_{im}+\sum_{i<j} \frac{e^2}{4\pi\epsilon_r}\frac{c^\dagger_{im}c_{im}c^\dagger_{jm}c_{jm}}{|\bm{r}_i-\bm{r}_j|} + \widetilde{V}.
\end{align}
The operator $c^\dagger_{im}$ creates an electron in band $m$ at site $i$. The first term in the Hamiltonian describes the dispersionless bands, while the second term represents the Coulomb interaction without exchange, justified by the large spatial separation between sites. This second term can be interpreted as a charge–charge interaction. Together, these two terms primarily determine the electronic band structure, with an associated energy scale ranging from several tens of meV to several eV. The final term, denoted as $\widetilde{V}$, arises from higher-order processes due to vacuum fluctuations. This term is non-Hermitian and has an energy scale on the order of 1 meV or smaller, making it significantly weaker than the first two contributions.
\begin{align}
    \widetilde{V}=\sum_{i,j}\xi_{ij} \hat{n}_i(c^\dagger_{j e}c_{j g}+c_{j e}c^\dagger_{j g})+V^{dd}_{ij}c^\dagger_{i e}c_{i g}c_{j e}c^\dagger_{j g},
\end{align}
where $g$ and $e$ denote the ground and excited band states, respectively. The first term represents the charge–dipole interaction, which is negligibly small in the absence of an external electric field due to energy and momentum conservation in vacuum. The second term corresponds to the dipole–dipole interaction between two artificial atoms. This interaction decays with distance as $1/R^3$. In the absence of an electric field, the artificial atoms do not possess permanent dipoles, but they can still interact via the exchange of excitations associated with the transition dipole moment. If one excited atom is located at the origin and a second ground atom at position $\bm{R}$, the latter experiences the dipole electric field of the former. The dipole–dipole interaction can be estimated as $V^{dd}=\frac{C_3}{R^3}(3(\bm{n}\cdot\bm{d}_1)(\bm{n}\cdot\bm{d}_2)-\bm{d}_1\cdot \bm{d}_2)$, where $\bm{n}=\bm{R}/R$. For a twist angle of 1.09°, $R=13.2$ nm and relative permittivity $\epsilon_r =4$. Taking $C_3$ is 1.44$\text{eV nm}^3/R^3\epsilon_r$, we obtain $V^{dd}\approx 1.25$ meV. 

Beyond the coherent hopping and interaction, decoherence is a critical factor in quantum systems. Possible sources of decoherence include thermal phonon noise, black-body radiation, and spontaneous emission. The estimated linewidths and lifetimes of the excited states are shown in Fig. 2c. At high temperatures, decoherence is dominated by thermal phonon noise. However, phonon effects can be effectively suppressed at low temperatures. Owing to the relatively weak electron–phonon coupling and the high Debye temperature of h-BN (approximately 600 K)~\cite{tohei2006debye}, the system is expected to maintain coherence at comparatively high operating temperatures. 
     
\begin{figure}
\centering
\includegraphics[width=1\linewidth]{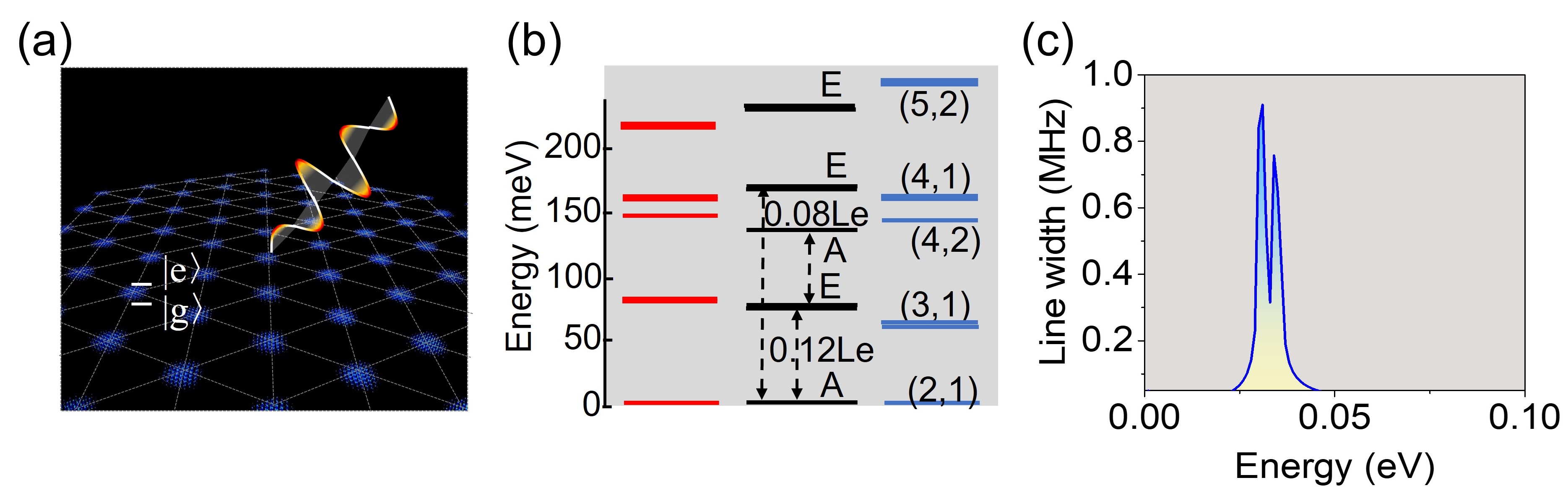}
\caption{Artificial-atom states in twisted bilayer h-BN. (a) DFT-calculated localized states on a triangular lattice. (b) Comparison of energy levels: left, results from the quantum-well model in Eq. 1; middle, DFT-calculated energy levels; right ones determined by quantum numbers. Thick and thin lines indicate twofold degenerate and nondegenerate levels, respectively. Labels A and E correspond to one-dimensional and two-dimensional irreducible representations, respectively. Dashed arrows indicate allowed optical transitions, and the numbers denote the corresponding transition dipole moments. (c) Calculated linewidth of the excited states.}
\label{fig:lifetime}
\end{figure}

\emph{Light-matter interface---}
Periodic artificial-atom states in twisted bilayer materials naturally emerge as a promising platform for engineering light–matter interactions and advancing various quantum technologies~\cite{bettles2016enhanced, ExponentialAsenjoGarcia2017, shahmoon2017cooperative, Manzoni2018, bekenstein2020quantum, Patti2021, rui2020subradiant, wei2021generation, srakaew2023subwavelength, Shah2024, Rubies-Bigorda2025Deterministic}. Moreover, the van der Waals interaction induced by the dipole–dipole interaction between two artificial atoms (see Eq. 3) leads to a blockade effect in otherwise dispersionless bands. The interaction energy scales as $\frac{1}{2\delta_F}(\frac{C_3}{R^3})^2 \approx \frac{2.07\text{eV}^2 \text{nm}^6}{2R^6\epsilon^2_r\delta_F}$. This provides a platform for quantum nonlinear optics~\cite{peyronel2012quantum}. For a twist angle of 1.09° in h-BN, assuming a typical detuning value ($\delta_F$) of 50 meV, the blockade shift is estimated to be as large as 0.001 meV (see Supporting Information for detailed calculations).

The full light-matter Hamiltonian is $H_{\text{tot}} = H_{\text{quad}} + U$:
\begin{align}
& H_{\text{quad}} = \sum_{i} \hbar\omega_{eg} b^\dagger_{i} b_{i}
  + \sum_{\Lambda=\pm} \int d\boldsymbol{k}\ \hbar c |\boldsymbol{k}| \mathcal{E}_{\Lambda}^{\dagger}(\boldsymbol{k}) \mathcal{E}_{\Lambda}(\boldsymbol{k}) \nonumber\\
&\quad + \sum_{i, \Lambda} \int d\boldsymbol{k}\ [g_{\bm{k},\Lambda}\exp(i\bm{k}\cdot\bm{R}_i) \mathcal{E}_{\Lambda}(\boldsymbol{k}) b^{\dagger}_{i} + \text{h.c.}].
\label{eqHquad}
\end{align}
Due to the extremely large blockade effect at the same site, the two-level artificial atoms are treated as hard-core bosons. $b_i^\dagger =|e_i \rangle \langle g_i |$ creates an atomic excitation at site $i$, satisfying the canonical commutation relations $[b_i, b^\dagger_j]=\delta_{ij}$. $  U  =  \sum_{i=1} u_i b^\dagger_i b^\dagger_i b_i b_i +\sum_{i,j} \frac{C_6}{|\bm{r}_i-\bm{r}_j|^6} b^\dagger_i b^\dagger_j b_i b_j,   u=\infty$ describes hard-core nonlinearity. The operator $\mathcal{E}_\Lambda^\dagger(\boldsymbol{k})$ creates a plane-wave photon with momentum $\bm{k}$ and polarization $\Lambda$. $g_{\bm{k},\Lambda}$ is the atom-photon coupling. 

\emph{Single-excitation physics--} 
The single-excitation Hamiltonian is identical to that in Eq. 4, which excludes the interaction terms. After tracing out the off-resonant vacuum electromagnetic modes, we obtain a non-Hermitian effective Hamiltonian given by~\cite{perczel2017photonic, williamson2020superatom}
\begin{align}
&H_{\text{eff}} =\sum_i(\hbar \omega_{eg}-i\frac{\Gamma_0}{2})b^\dagger_i b_i \nonumber \\
&+ \frac{3\pi\hbar\Gamma_0c}{\omega_{eg}}\sum_{i\ne j, \alpha, \beta}  G_{\alpha, \beta}(\bm{r}_i-\bm{r_j})b^\dagger_i b_j.
\label{eq:Hp_bright_rev}
\end{align}
Here, $\Gamma_0 = \frac{d^2\omega^3_{eg}}{3\pi \varepsilon_0 \hbar c^3}$ is the radiative linewidth of a single artificial atom in free space. The tensor $G(\bm{r})$ denotes the dyadic Green’s function, which describes the dipole–dipole (spin–spin) interactions between two artificial atoms~\cite{Manzoni_2018, PhysRevA.95.033818}. $\alpha$ and $\beta$ label the polarization components of the artificial atoms. In a periodic, infinite array of artificial atoms, it obeys Bloch’s theorem. Thus, the single-excitation Hamiltonian in momentum space can be written as $H_{\text{eff}} = \Delta(\boldsymbol{p}) b^\dagger_{\boldsymbol{p}} b_{\boldsymbol{p}}$, where the bosonic operator $b^\dagger_{\bm{p}}$ creates an excitation of collective atomic spin waves between $|g \rangle$ and  $|e\rangle$. The real and imaginary parts of $\Delta(\bm{p})$, denoted by $\Delta_{\bm{p}}$ and $\Gamma_{\bm{p}}$, correspond to the frequency shift of the mode (relative to the bare atomic frequency $\omega_{eg}$) and its associated decay rate, respectively. 

\begin{figure}
\centering
\includegraphics[width=1.0\linewidth]{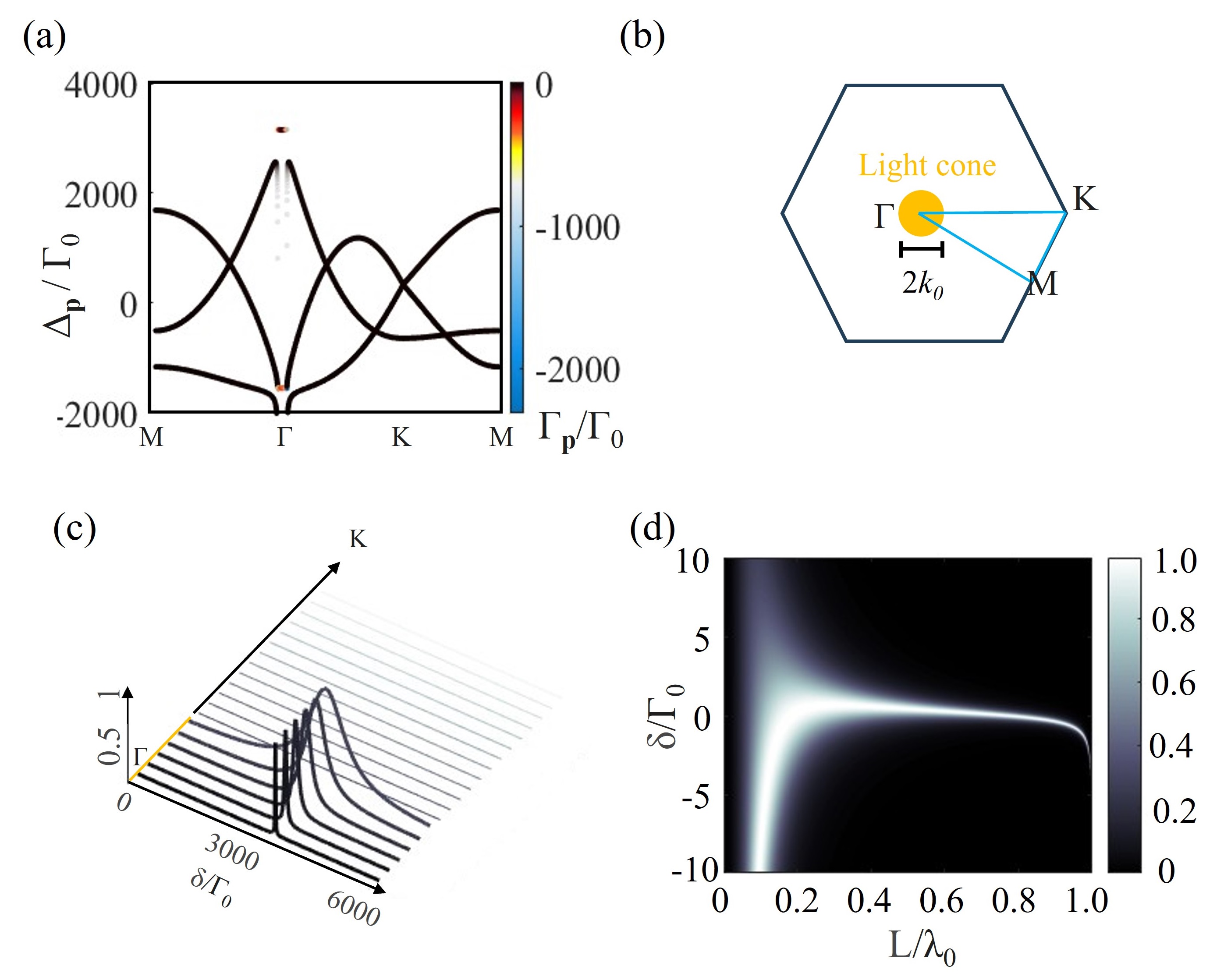}
\caption{(a) The Lamb shift and decay rate in a triangular moiré superlattice. (b) Brillouin zone of the triangular lattice, with the light cone indicated by yellow. (c) Calculated reflection along momentum line from $\Gamma$ to K. (d) Calculated reflection for different detunings and lattice constants (or twist angle).}
\label{fig:placeholder}
\end{figure}

The calculated $\Delta(\bm{p})$ for a triangular lattice with lattice constant $d=0.02\lambda_0$ is shown in Fig.~\ref{fig:placeholder}(a). Based on the value of $\Gamma_{\bm{p}}$, we find that the superradiance occurs only inside the light cone, whereas the subradiance occurs outside it. Here the light cone is defined as $|\bm{p}| < k_0$ with $k_0 = \omega_{eg} / c$, as schematically illustrated in Fig. 3(b). Since the lattice spacing $d<\lambda_0/2$, the light cone lies entirely within the first Brillouin zone. In moiré superlattices, the inter-well spacing can be tuned via the twist angle and typically ranges from 1/10000 to 1/10 $\lambda_0$~\cite{hong2026highly}. Consequently, the interaction between two artificial atoms is significantly stronger than that in arrays of alkali atoms, leading to correspondingly larger Lamb shifts and decay rates.

After incorporating the probing light to detect the optical properties of this artificial atomic array, the single-excitation Hamiltonian in momentum space can be written as follows\cite{wang2025universal}:
\begin{align}
  & H_{\bm{p}} =  \Delta(\boldsymbol{p}) b^\dagger_{\boldsymbol{p}} b_{\boldsymbol{p}} + \int_{-\infty}^{\infty} \!\!\! dk_z \, E_{\bm{p}}(k_z) C_{\bm{p}}^{\dagger}(k_z) C_{\boldsymbol{p}}(k_z) \nonumber \\
  &+ \int_{-\infty}^{\infty} \!\!\! dk_z \, \left[ g_{\boldsymbol{p}}(k_z) C_{\boldsymbol{p}}^{\dagger}(k_z) b_{\boldsymbol{p}} + \text{h.c.} \right] \; (\text{for } |\bm{p}| < k_0).
\label{eq:Hp_bright_rev_1}
\end{align}
Here, $C^\dagger_{\bm p}(k_z)$ is the creation operator of the probing light, where $k_z$ denotes the momentum component orthogonal to the 2D array. $E_{\bm{p}}(k_z) = c\sqrt{|\bm{p}|^2 + k_z^2}$ is the energy of a probing photon, and $g_{\boldsymbol{p}}(k_z)$ is the coupling strength between the artificial atom and the probing light. 

According to the input-output theory, the transmission $t(\delta)$ and reflection $r(\delta)$ coefficients for a plane-wave photon incident from one side of the array with in-plane momentum $\bm{p}$ and normal component $k_z>0$ are given by~\cite{shahmoon2017cooperative, Wang2025MultiExcitation}

\eq{
   r(\delta)=&\frac{-i\Gamma(\bm{p})/2}{i\Gamma(\bm{p})/2 + \delta-\Delta_{\bm{p}}}\\
   t(\delta)=&1+r(\delta)}
where the detuning is defined as $\delta=E_{\bm{p}}(k_z)-\omega_{eg}$. The transmitted photon retains the same momentum $(\bm{p}, k_z)^T$ as the incoming photon, while the reflected photon carries momentum $(\bm{p}, -k_z)^T$. The corresponding transmission and reflection probabilities are $|t|^2$ and $|r|^2$, respectively. The calculated reflection spectrum is shown in Fig.~\ref{fig:placeholder}(c). Inside the light cone ($|\bm{p}| < k_0$), the collective excitation (bright spin wave) decays away from the lattice in the transverse direction due to the its finite lifetime $1 / \Gamma(\bm{p})$. Outside the light cone ($|\bm{p}| > k_0$), the collective (dark) spin wave has an infinite lifetime with zero decay. Thus, the signature of superradiance or subradiance significantly influences the optical properties. Beside, inside the light cone, by carefully tuning the detuning, one can achieve either nearly perfect reflection or transmission with $|r|^2 \approx$ 100$\%$ or 0, repectively.

Based on the optical properties described above, moiré superlattices can act as tunable mirrors for photons, switch between long-lived dark spin waves and short-lived bright spin waves~\cite{Manzoni2018, ExponentialAsenjoGarcia2017,shahmoon2017cooperative, bettles2016enhanced, manzoni2018optimization}, when appropriate twist angles and wavelengths are chosen~\cite{marangos1998electromagnetically}, as shown in Fig. 3(d). The dark spin waves that enable photon storage can also be realized via local control of the detuning of the energy levels. In contrast to alkali-atom systems, detuning in moiré superlattices can be readily achieved using electrostatic gates or local electric fields. This capability paves the way for gate-controlled single-photon quantum optics.

\emph{Two-excitation nonlinearity---}
We now return to Eq. 4 and focus on the blockade interaction. An artificial atomic array in moiré superlatices also serves as a promising platform for generating strong optical nonlinearities. Here the intrinsic nonlinearity of two-level atoms, described by the interaction term $U$, is to induce the optical nonlinearity~\cite{williamson2020superatom}. When a weak coherent beam with photon rate $N_p$ (photons per unit time) illuminates the system, the probability of inelastic photon scattering is $\sim N_p \Gamma \frac{r^2_d}{A}$  for two-level nonlinearities, where A is the area of the beam waist area. The effective blockade radius $r_d$ is estimated as 200 nm at 10K in 1.09°-twisted h-BN. 

The blockade area $r^2_d$ is close to $A$, and thus the nonlinear interactions between atoms in the $|e\rangle$ state are strong. Consequently, the system supports strong nonlinear photon scattering, even without requiring additional coupling to Rydberg levels. The Rydberg electromagnetically induced transparency (EIT) scheme commonly used in neutral-atom systems~\cite{eisaman2005electromagnetically} becomes unnecessary in moiré superlatices, making quantum nonlinearities easier to realize. Systems exhibiting strong photon–photon interactions enable a number of unique applications, including quantum-by-quantum control of light fields, single-photon switches and transistors, all-optical deterministic quantum logic, quantum gates and the realization of strongly correlated states of light and matter, quantum simulator or quantum repeater~\cite{chang2014quantum, moreno2021quantum, williamson2020superatom, cidrim2020photon, liu2025millisecond}.


\emph{Material database---} Dispersionless bands and atom-like localized states are ubiquitous in a wide range of twisted materials at small twist angles. Examples include twisted bilayer $PbS$, $SrTiO_3$, $In_2Se_3$ and $Bi_2Se_3$. Twisted bilayer $PbS$ and $SrTiO_3$ form square moiré lattices, where the energy separation between dispersionless bands can reach several hundred meV~\cite{shahed2025prediction, song2025twisted}. In contrast, twisted bilayer $MoS_2$, $In_2Se_3$ and $Bi_2Se_3$ exhibit triangular lattices, with much smaller energy separations—typically on the order of a few meV at small twist angles~\cite{song2025twisted, li2021lattice, zhao2020formation}. Experimentally, a variety of twisted materials have already been realized, including $PbS$, $SrTiO_3$ and $MoS_2$~\cite{wang2022strong, kim2025twisted, sha2024polar}. Moreover, the twist angle can be tuned to extremely small values, reaching as low as 0.01° in systems such as twisted h-BN and $MoS_2$~\cite{hong2026highly,quan2021phonon}.

Based on our proposal, the key requirements for twisted layers to serve as suitable quantum-optical systems are as follows: (1) the presence of at least two dispersionless bands (neglecting spin), (2) deep and well-isolated dispersionless bands within the band gap to minimize hybridization with other bands, and (3) symmetry-allowed optical transitions. Since direct electronic structure calculations for all moiré superlattices are computationally infeasible, these key properties can instead be inferred from the electronic structure of the constituent monolayers. Materials with large band gaps and finite effective masses in the conduction or valence bands tend to produce dispersionless bands at small twist angles. Furthermore, interlayer coupling typically leads to sizable energy separations between these dispersionless bands. Importantly, this energy separation determines the operational wavelength for quantum-optical applications.

Symmetry plays a central role in determining optical excitations in twisted bilayer systems. The moiré point group can be inferred from that of the constituent layers, assuming the two layers are stacked without inversion and twisted by a relative angle. The formation of in-plane strain vortices and out-of-plane strain bulges generically breaks mirror and inversion symmetries while preserving rotational symmetries, so that the resulting moiré symmetry is governed primarily by the rotational symmetry of the  constituent layers. Consequently, low-symmetry groups ($C_1, C_i$, and $C_s$ as well as $S_2$ reduce to $C_1$, twofold rotational groups ($C_2, C_{2h}, C_{2v}$ and $S_4$) to $C_2$, threefold groups ($C_3, C_{3v}$, $C_{3h}$, $S_6$) to $C_3$. $D_2, D_{2h}$ and $D_{2d}$ to $D_2$. $D_3$, $D_{3d}$ and $D_{3h}$ reduce to a $D_3$ moiré point group after twisting. Layers with $C_4, C_{4h}$ and $C_{4v}$ produce $C_4$ symmetry, while $D_4$, and $D_{4h}$ lead to $D_4$ moiré point group. Likewise, $C_6, C_{6h}$, and $C_{6v}$ yield $C_6$, whereas $D_6, D_{6v}$ and $D_{6h}$ result in $D_6$ symmetry after twisting. Based on these symmetry considerations, we compile a database of 741 candidate materials (excluding oxides due to defect susceptibility), characterized by their interlayer coupling, band gaps, and chemical compositions~\cite{campi2023expansion, mounet2018two}. The resulting energy separations span from a few meV to several eV, depending on the material and twist angle.

\emph{Discussion.}  Compared with other systems, especially alkali atomic arrays, moiré systems offer several advantages. The localized states form a self-organized two-dimensional array, making the emitters naturally scalable. The quantum-well structures are identical if materials are carefully fabricated to minimize strain. The distance between any two artificial-atom states is tunable via the twist angle, allowing adjustable spacing and interactions. Moreover, semiconductor techniques, such as electrostatic gates, can be readily used to facilitate optical control. Finally, the extensive materials database of moiré superlattices allows coverage across a wide range of light wavelengths.


\emph{Acknowledgement.~}

\appendix

\bibliographystyle{apsrev4-2}
\bibliography{library}

\end{document}